\documentclass{eas}
\usepackage{graphicx}
\begin{document}

\title{ {\underline {\bf MIMAC}}: A micro-tpc matrix for directional detection of dark matter}

\author{D.~Santos} 
\address{  LPSC, Universite Joseph Fourier Grenoble 1, CNRS/IN2P3, Institut Polytechnique de Grenoble}
\author{J.~Billard}\sameaddress{1}

\author{  G.~Bosson} \sameaddress{1}
\author{J.L. ~Bouly}\sameaddress{1}
\author{ O.~Bourrion}\sameaddress{1}
\author{ Ch. ~Fourel}\sameaddress{1}
\author{ O.~Guillaudin}\sameaddress{1}
\author{ F.~Mayet}\sameaddress{1}
\author{ J.P.~Richer}\sameaddress{1}
\author{A. Delbart}
\address{	IRFU,CEA Saclay, 91191 Gif-sur-Yvette cedex}
\author{E.~Ferrer}\sameaddress{2}
\author{ I.~Giomataris}\sameaddress{2}
\author{ F.J.~Iguaz}\sameaddress{2}
\author{ J.P.~Mols}\sameaddress{2}
\author{C.~Golabek}
\address{	LMDN, IRSN Cadarache, 13115 Saint-Paul-Lez-Durance} 
\author{ L.~Lebreton}\sameaddress{3}


%
%
\begin{abstract}

Directional detection of non-baryonic Dark Matter is a promising search strategy for discriminating  WIMP events from background ones. This strategy requires both a  measurement of the recoil energy down to a few keV and 3D reconstruction of tracks down to a few mm. The MIMAC project, based on a  micro-TPC matrix, filled with $\rm CF_4$ and $\rm CHF_{3}$ is being developed. The first results of a chamber prototype of this matrix,  on low energy nuclear recoils ($\rm ^1H$ and $\rm ^{19}F $) obtained with  mono-energetic neutron fields are presented. The discovery potential of this search strategy is  illustrated by a realistic case accessible to MIMAC.
\end{abstract}
\maketitle
\section{Introduction}

Directional detection of Dark Matter is based on the fact that the solar system moves with respect to the center of our galaxy with a mean velocity of roughly 220 km/s (\cite{spergel}). Taking into account the hypothesis of the existence of a galactic halo of DM formed by WIMPs (Weakly Interacting Particles) with a negligible rotation velocity, we can expect a privileged direction for the nuclear recoils in our detector, induced by elastic collision with those WIMPs.

The MIMAC (MIcro-tpc MAtrix of Chambers) detector project tries to detect these elusive events by a double detection: ionization and track, at low gas pressure with low mass target nuclei (H, $^{19}$F or $^3$He). In order to have a significant cross section we explore the axial, spin dependent, interaction on odd nuclei. The very weak correlation between the neutralino-nucleon scalar cross section and the axial one, as it was shown in 
(\cite{PLB}) (\cite{Vasquez}), makes this research complementary to the massive target experiments.

\section{MIMAC prototype}

The MIMAC prototype consists of one of the chambers of a large matrix  detector allowing to show the ionization and track measurement performance needed to achieve the directional detection strategy.
The primary electron-ion pairs produced by a nuclear recoil in one chamber of the matrix are detected by driving the electrons to the grid of a bulk micromegas (\cite{bulk}) and producing the avalanche in a very thin gap (128 or 256$\mu$m) adapted to the gas pressure. 

\begin{figure}[h!]
\begin{center}
\includegraphics[scale=0.5]{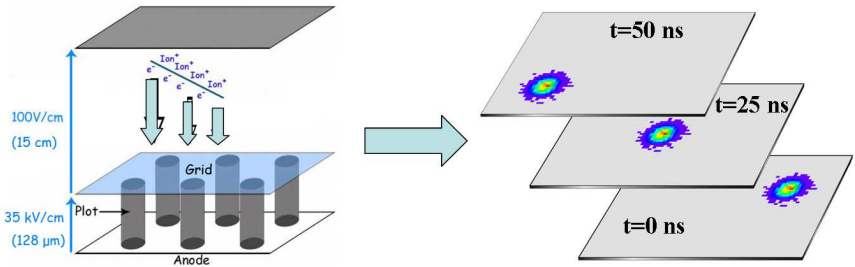}
\caption{Track reconstruction in MIMAC. The anode was read every 25 ns in the first version, now every 20 ns, giving  the 3D track  envelope from the consecutive number of images, defining the event. The drift velocity of the primary electrons, needed to recover the third dimension length,  must be determined by an independent dedicated experiment.}
\label{recon}
\end{center}
\end{figure}

\begin{figure}[h!]
\begin{center}
\begin{minipage}[h!]{5.3 in}
\includegraphics[scale=0.33]{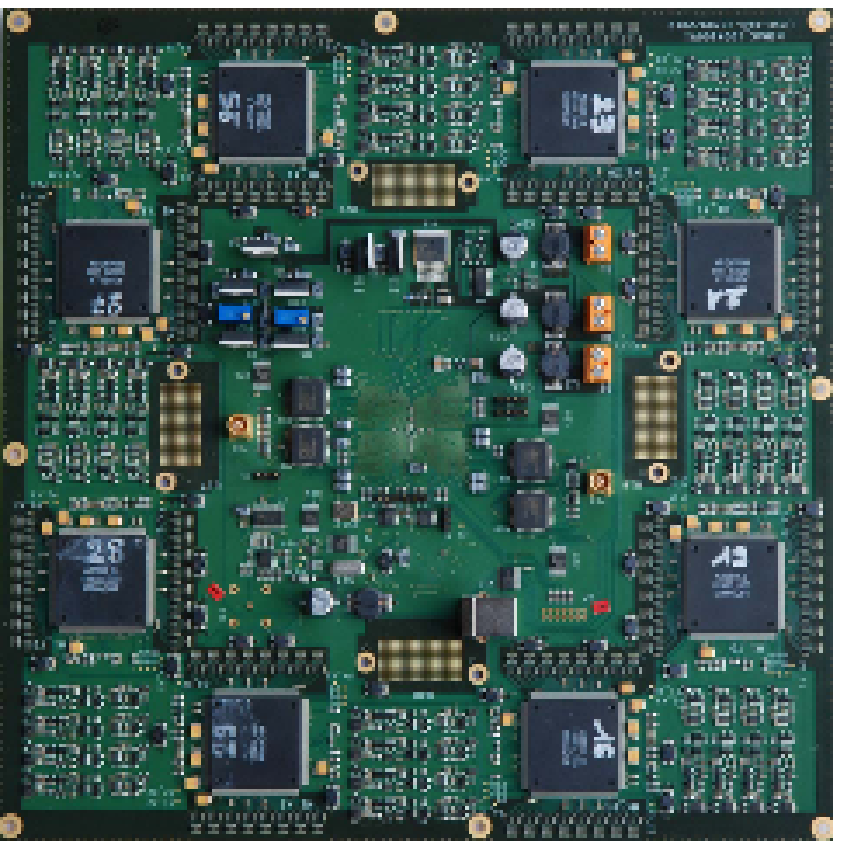} \hspace{.2in}
\includegraphics[scale=0.45]{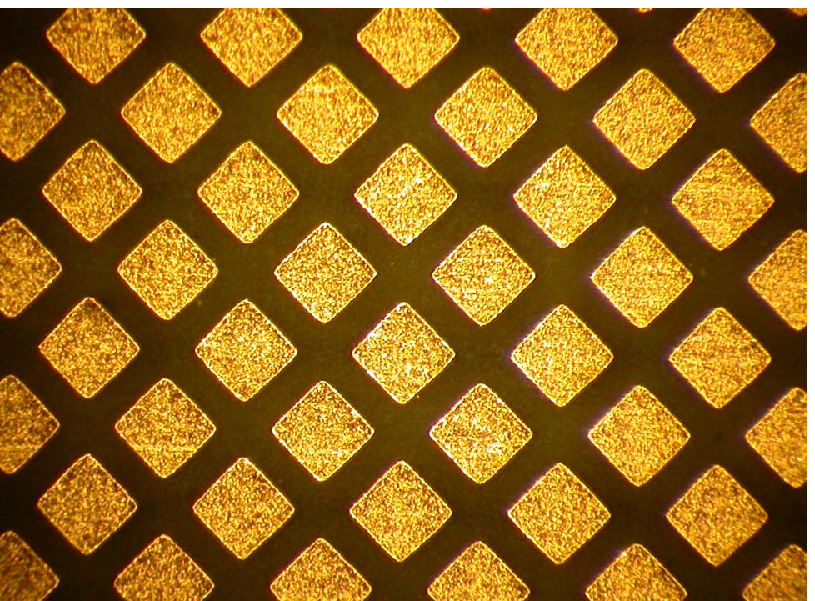} \hspace{.2in}
\includegraphics[scale=0.55]{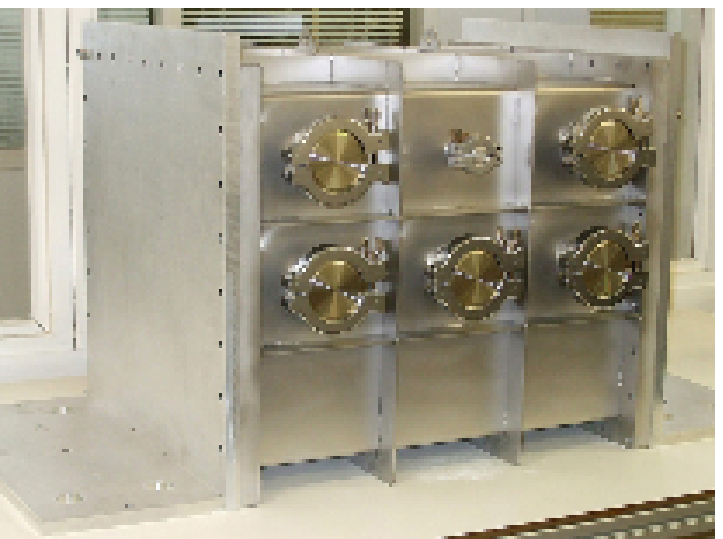}
 \caption{ : (Left): MIMAC electronic card with the 8 chips (64 channels each) reading-out the 512 channels of the 10cmx10cm MIMAC prototype.(Center): A small portion of the 10cmx10cm  pixelized anode. (Right): The bi-chamber module to be installed at Modane in March 2012.}
\label{proto}
\end{minipage}
\end{center}
\end{figure}

As schematically shown in figure  \ref{recon}, the electrons are moving towards the mesh in the drift space and are projected on the anode thus allowing to get 
information on X and Y coordinates of the recoil track.
To determine the X and Y coordinates with a 100 $\mu$m spatial resolution, a bulk micromegas  with a 10 x 10 cm$^2$ active area, segmented in pixels with a pitch of 350 $\mu$m was used, shown in fig. 2 (center).
 In order to reconstruct the third coordinate Z of the recoil track, the LPSC  has developed a self-triggered electronics able to perform the anode sampling at a frequency of 50 MHz.
This includes a specially developped 64 channel chip (\cite{richer}) in its second version,  associated to a DAQ  (\cite{bourrion}). The electronic card including 8 chips coupled to the anode is shown in fig. \ref{proto} (left). In total 512 channels are read every 20 ns.

The 3D tracks are obtained from consecutive read-outs of the anode, every 20 ns, defining the event. To get the length and the orientation of the track, an independent measurement of the drift velocity is needed. 
An important feature concerns the adaptation of the drift velocity value  in the gas mixture used to the sampling  time of the anode. Pure  $ CF_{4}$ is too fast to allow to have enough 20 ns slices  read-out defining a small  recoil track (a few mm). We have determined the proper mixture to be used in the MIMAC detector coping with very small recoil track lengths at low recoil energies (5- 50 keV).  We have experimentally demonstrated that with a percentage of 30 to 50\% of $CHF_{3}$, depending on the energy range of the recoils,  we can slowdown the drift velocity of the electrons in the drift space,  allowing to have at least 2 or 3 read-outs of the anode to define the direction of the track. Fig. 3 shows the Magboltz calculation (\cite{Magb}) of the drift velocity in   $ CF_{4}$ as a function of the percentage of  $CHF_{3}$. Preliminary measurements of the drift velocity, using  an alpha particle source,  have been performed  in different gases in the MIMAC prototype  fitting well with Magboltz simulations.

\begin{figure}[h!]
\begin{center}
\begin{minipage}[h!]{5.3 in}
\includegraphics[scale=0.28]{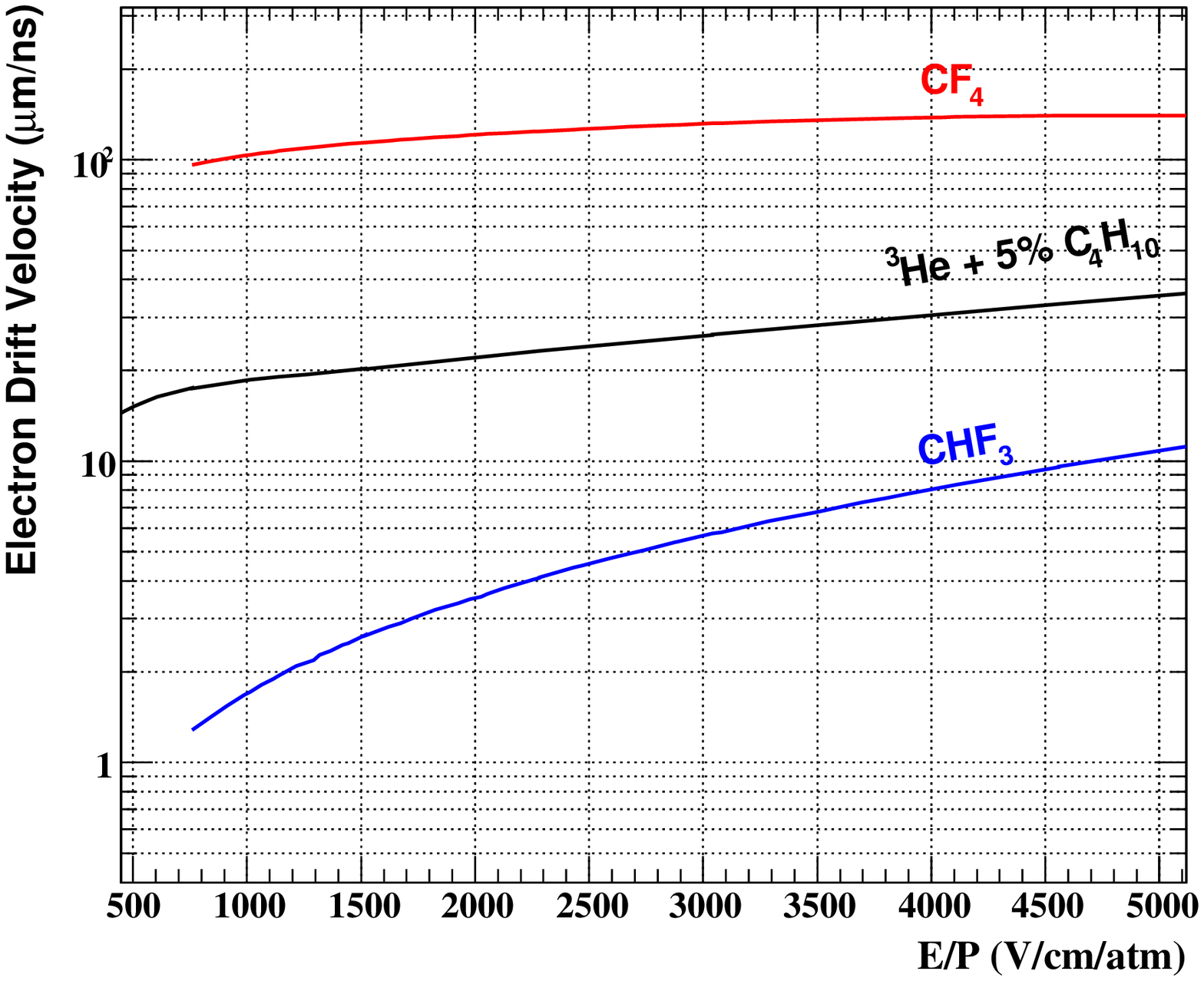}\hspace{.2in}
\includegraphics[scale=0.28]{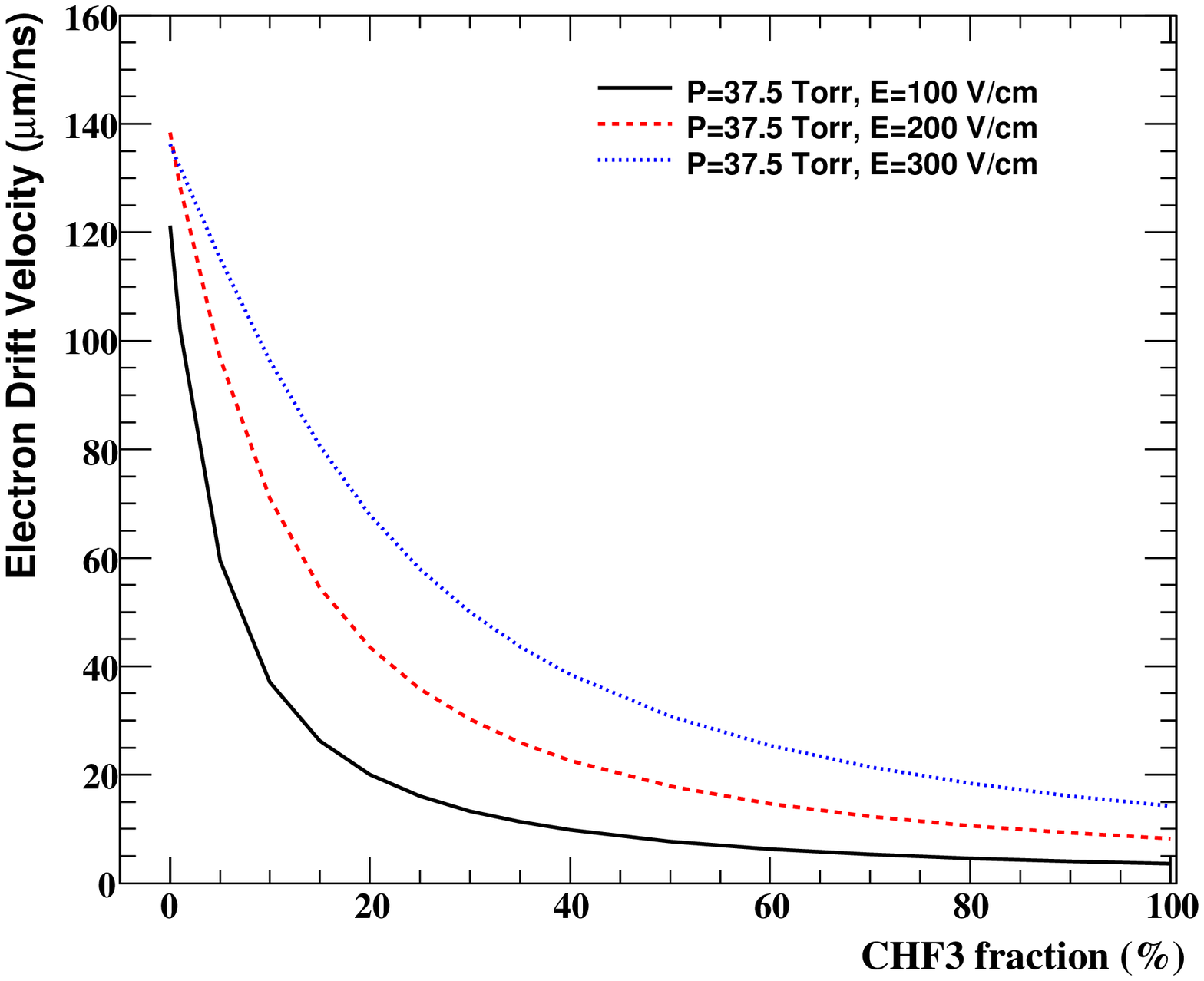}
\caption{Left: Electron drift velocity as a function of the electric field over the pressure in the chamber calculated with the Magboltz code in pure $ CF_{4}$ , in pure  $ CHF_{3}$ and in$^4$He + 5$\% \rm C_4H_{10}$ mixture for comparison. Right: The electron drift velocity in a mixed of $ CF_{4}$  and  $ CHF_{3}$ in 50 mbar as a function of the percentage of the last one  for three different drift electric fields }
\label{CHF3}
\end{minipage}
\end{center}
\end{figure}

To determine  the  kinetic recoil energy of a recoil event we need to know the ionization quenching factor (IQF) of such nuclear recoil in the gas mixture. We have developed at the LPSC a dedicated experimental facility to measure such IQF. A description of such facility and a precise assessment of the available ionization energy by a  measurement of the IQF has been performed by  (\cite{santos.q}) in $^4$He + 5$\% \rm C_4H_{10}$ mixture within the dark matter  direct detection energy range (between 1 and 50 keV) .
The IQF  measurents of  $^{19}F $ in $ CF_{4}$ has been performed recently (\cite{OG}) giving the recoil energy threshold to be expected in  directional detection using a $^{19}F $ target.  We have detected  $^{19}F $ with 5 keV kinetic energy leaving in ionization only 1.2 keV. A publication concerning these important measurements for dark matter direct detection, as a funcion of the gas pressure and up to 50 keV,  is in preparation.

\section{ Experimental results}

In order to validate the MIMAC prototype we had to be able to detect simultaneouly the energy and the track of nuclear recoils in the range of 1- 100 keV.  Neutrons are the ideal tool to produce nuclear recoils by elastic collision as the Wimp event signal searched. We have performed dedicated experiments at AMANDE facility (\cite{Lena}) (IRSN - Cadarache (FRANCE)) producing neutron monochromatic fields from (p,n) reactions. The neutron energy is selected from the maximum energy produced at the resonance at 0$^o$ by a moving arm around the target. In such a way, neutron energies covering  10 to  565 keV are available. Two different neutron fields were used at 144 keV and 565 keV maximum energies on C$_4$H$_10$ and CF$_4$ respectively. 
A X-ray calibration has been produced from $^{109}$Cd and the $^{55}$Fe radioactive sources giving the  3.05 keV and  5.9 keV lines respectively. The X-ray spectrum is shown in fig. 4 for a high gain run. This spectrum is the projection of all the events producing tracks in the active volume. 

\begin{figure}[h!]
\begin{center}
\includegraphics[scale=0.25]{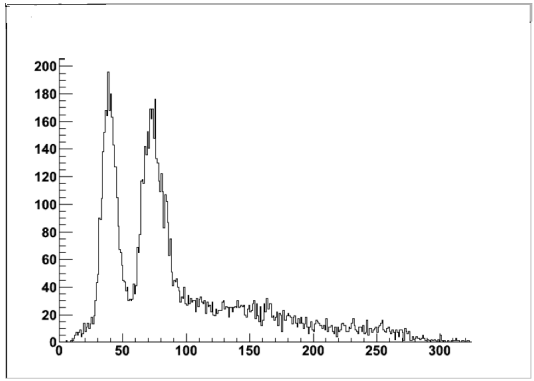}
\caption{ X-ray calibration with $^{109}Cd$  and $^{55}Fe$ sources. We can see the 3.05 keV and the 5.9 keV lines. The second one broadened by the $K_{\beta}$  6.4 keV not resolved.}
\label{calib}
\end{center}
\end{figure}

A very important point for dark matter detection is the ability to separate gamma ray background from nuclear recoils. For a given energy, an electron track  is an order of magnitude longer than a recoil one. In such a way,  the discrimination of electrons from nuclei recoils by using both ionization energy and track length information, is performed in its first step.  The plot of the length of the tracks versus their ionization energy, shown in fig.5(left) was the first raw analysis during the experiment with 144 keV neutrons on C$_4$H$_{10}$,  showing the high quality of observables available from the read-out. This discrimination is even improved at low energies  event by event by a new degree of freedom called the normalized integrated straggling (NIS) of the track, defined as the sum of all the angular deviations along the track, normalized by the ionization energy of the particle. 
The NIS histogram shown is produced using the regions of the length vs. energy plane where no confusion arises between electrons and recoils giving for discrimination a clear cut on the NIS value.
This NIS discrimination degree of freedom  is  available from  linking  the barycentres of each 20ns slice read-out, putting in evidence the difference in the collision straggling between electrons and recoils. The track of recoils are more straight than those of the electrons due to their larger mass, as is illustrated in fig 5 (right). 

\begin{figure}[h!]
\begin{center}
\begin{minipage}[h!]{5. in}
\includegraphics[scale=0.8]{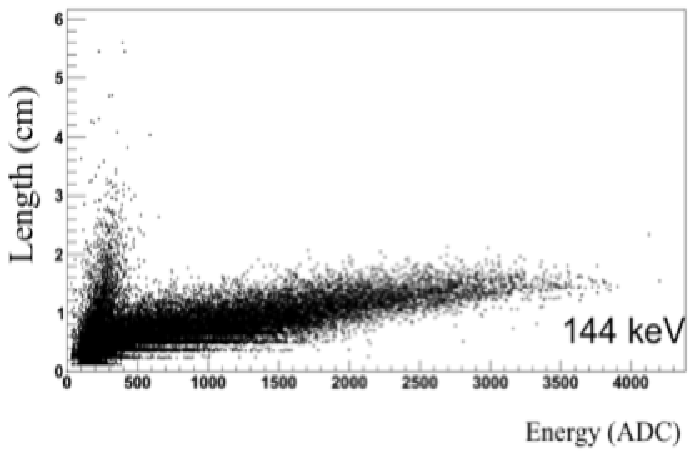} \hspace{.1in}
\includegraphics[scale=0.29]{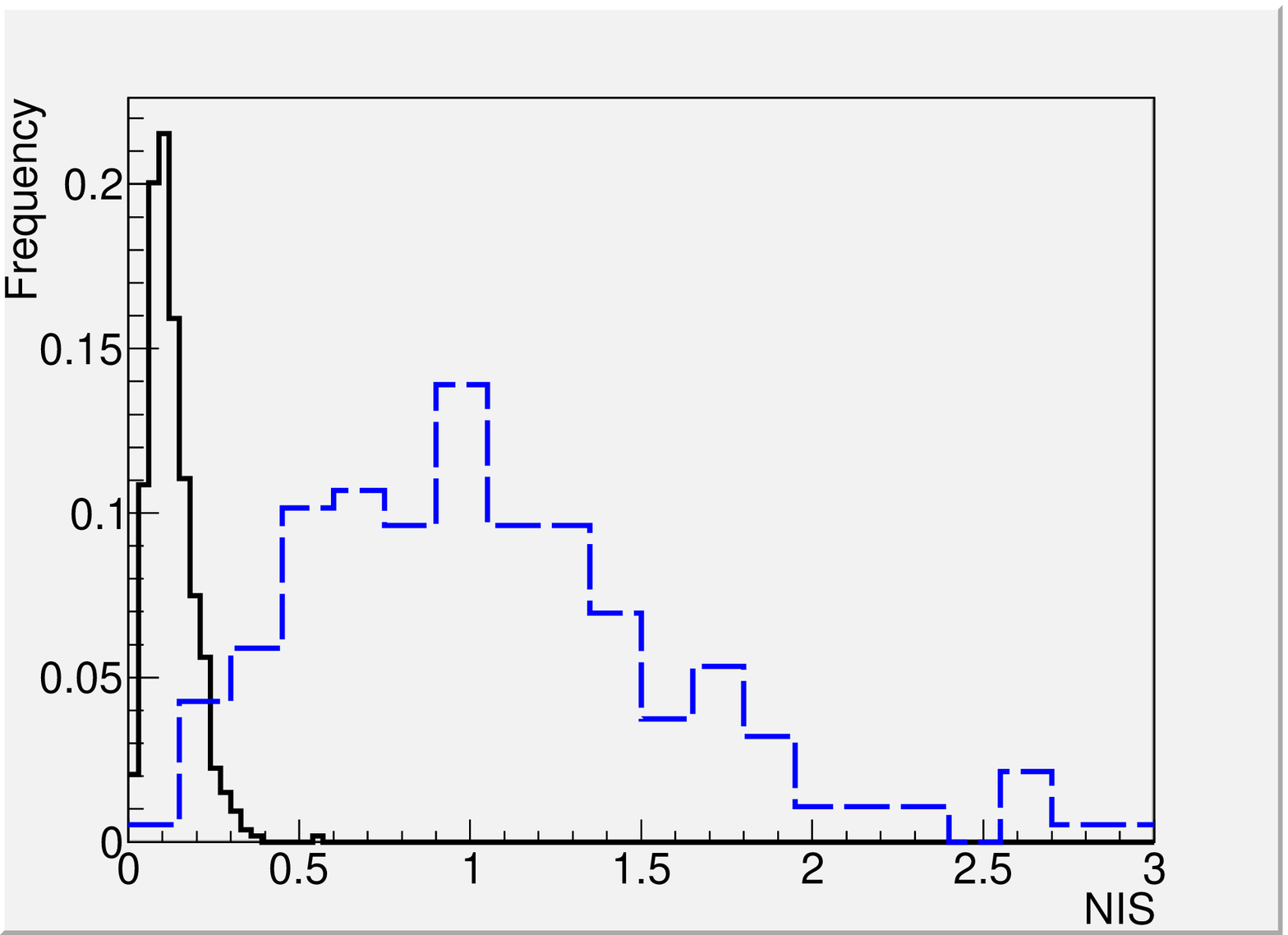} 
 \caption{ : (Left):  The length vs. ionization energy of all the raw events produced by  the neutrons (144 keV)  on $C_4H_10$ and gamma rays produced on the target. (Right): The NIS, obtained as a sum of all the step deviations (every 20 ns) normalized by the energy, in black for recoils and in blue for electrons.}
\label{e-r_discr}
\end{minipage}
\end{center}
\end{figure}

\begin{figure}[h!]
\begin{center}
\includegraphics[scale=0.27]{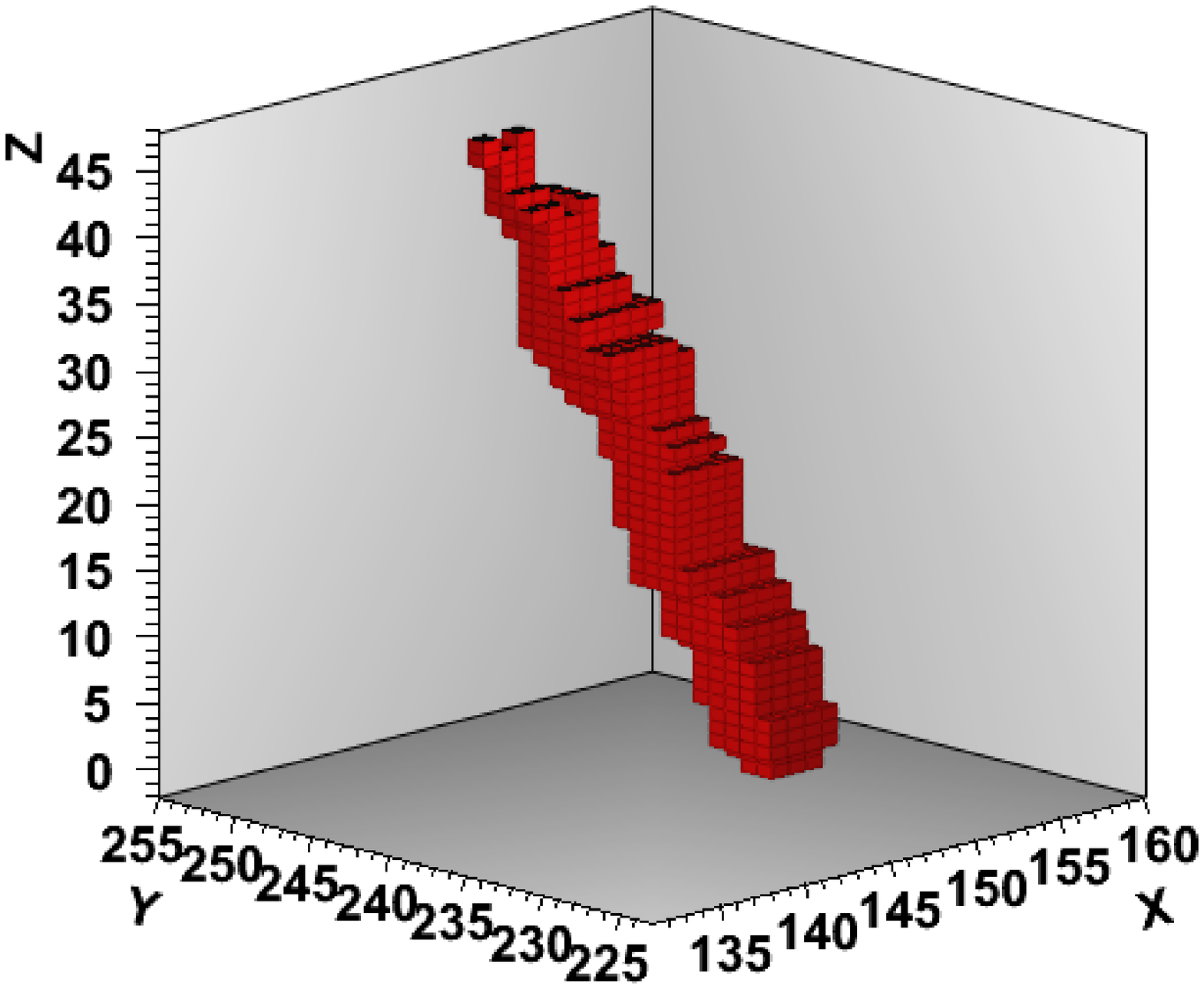}\hspace{.15in}
\includegraphics[scale=0.51]{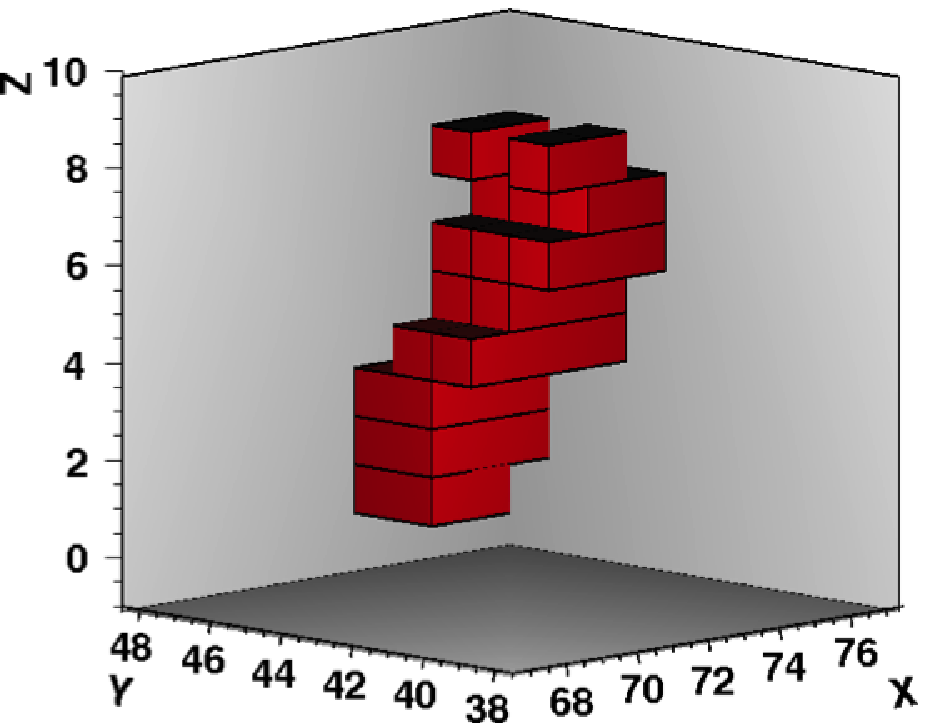}
\caption{ Left: A 57 keV 3D proton track in  $ CF_{4}$  plus 30\%  $ CHF_{3}$ in 50 mbar. Right: A 40 keVee fluorine track in the same gas mixture}
\label{proton3D&fluorine3D}
\end{center}
\end{figure}

In order to understand the wealth of the information obtained by the read-out, we show in fig. 6 the 3D reconstruction of a proton and fluorine tracks showing the shape of each X-Y  consecutive read-outs. A straightforward idea of the direction of the track is yielded by the barycentre of each patch. The shape of each patch gives information on the direction of tthe track inside the 20 ns drift, that is extracted by a maximization of a likelihood in the final step of the analysis (\cite{Billard.cygnus}) giving at the same time an estimation of the distance from the anode of the track, taking into account the longitudinal and transverse difusion of the primary electrons opening the possibility of a fiducialization of the active volume.

\section{ MIMAC prospective}

The MIMAC first bi-chamber prototype consisting of two chambers of 10x10x25 cm$^3$, shown in fig. 2 (right), will be installed at the Laboratoire Souterrain de Modane (LSM-France) in March 2012.  We are working on the 20x20 cm$^2$  pixelized micromegas having 1024 channels of read-out. The demonstrator of 1m$^3$ should follow the validation of the first bi-chamber  prototype. The 1m$^3$ will be built as the reproduction, 100 times, of the bi-chamber (20x20x25 cm$^3$) module, divided in two big chambers 0.5 m$^3$ active volume each.
This should be follow by a large matrix detector  50 m$^3$. We have shown recently that the directional detection opens the way for a discovery with a reasonable number of WIMP events (\cite{billard.disco}).
Figure 7  presents, in the proton axial cross-section versus WIMP mass plane, the discovery region and the background free exclusion limit for a  50 m$^3$ during 3 years. The  1m$^3$ demonstrator detector may allow for a discovery down to 10$^{-2}$ pb and for an exclusion down to 4. 10$^{-3}$ pb.

\begin{figure}[h!]
\begin{center}
\includegraphics[scale=0.45]{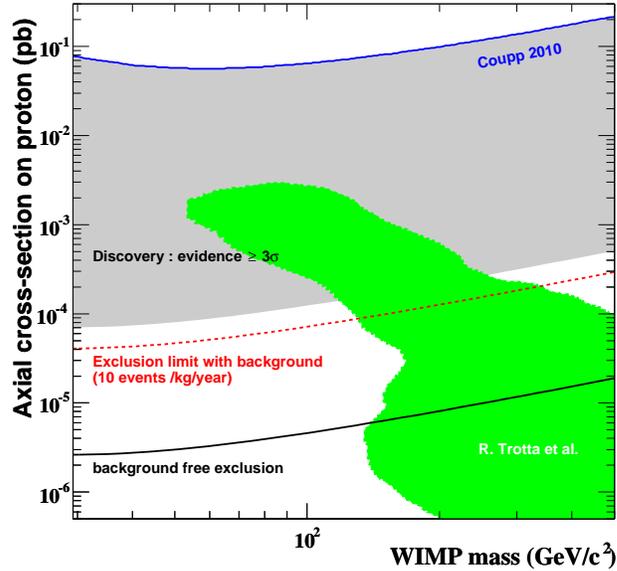}
\caption{Axial cross section on proton (pb) as a function of the WIMP mass. The black curve shows the 3$\sigma$ discovery region. The shaded region shows the difference with respect to the best exclusion curve today (\cite{coupp}) in the spin-spin interaction vs. WIMP mass plane. The exclusion curves available at such exposure are shown in red. The red dashed curve shows the exclusion curve with 10  background events /( kg  year) contamination. The green region represents the SUSY models calculated by (\cite{trotta}).}
\label{ex&disc}
\end{center}
\end{figure}


\end{document}